%% file: main.tex
\documentclass[]{acmart}

\AtBeginDocument{%
  \providecommand\BibTeX{{%
    \normalfont B\kern-0.5em{\scshape i\kern-0.25em b}\kern-0.8em\TeX}}}
    
\setcopyright{acmcopyright}
\copyrightyear{2022}
\acmYear{2022}
\acmDOI{10.1145/1122445.1122456}

\usepackage{tikz}
\usepackage{amsmath}
\usepackage{graphicx}
\usepackage{multicol}
\usepackage{subcaption}
\usepackage{tabularx}
\usepackage{balance}
\usepackage[subtle]{savetrees}

\usepackage{url}

\hypersetup{breaklinks=true}

\newcommand{\blind}[2]{#2}

\usepackage{natbib} 
\newenvironment{myquote}%
  {\list{}{\leftmargin=0.05in\rightmargin=0.05in}\item[]}%
  {\endlist}

\begin{document}

\date{}

\title[To Patch, or not To Patch? That is the Question]{To Patch, or not To Patch? That is the Question: A Case Study of System Administrators' Online Collaborative Behaviour}

\author{Adam Jenkins}
\email{adam.jenkins@ed.ac.uk}
\orcid{0001-7865-0087}
\affiliation{%
  \institution{University of Edinburgh}
  \city{Edinburgh}
  \country{United Kingdom}
}

\author{Maria Wolters}
\email{maria.wolters@ed.ac.uk}
\affiliation{%
  \institution{University of Edinburgh}
  \city{Edinburgh}
  \country{Scotland}
  }

\author{Kami Vaniea}
  \email{kvaniea@inf.ed.ac.uk}
\affiliation{%
  \institution{University of Edinburgh}
  \city{Edinburgh}
  \country{United Kingdom}
}

\renewcommand{\shortauthors}{Jenkins et al.}

\begin{abstract}

\input{Text/Abstract.tex}
\end{abstract}



\maketitle

\section{Introduction}
\vspace{-1mm}
\label{sec:introduction}

\input{Text/Introduction}

\section{Related Work}
\vspace{-1mm}
\label{sec:realtedWork}
\input{Text/Literature.tex}

\section{Methodology}
\vspace{-1mm}
\label{sec:methods}
\input{Text/methods.tex}

\section{The Cast: Places \& Players}
\vspace{-1mm}
\label{sec:places+players}
\input{Text/PlacesPlayers}

\section{The Case of  KB4034679 \& KB4034664}
\vspace{-1mm}
\label{sec:story}

\input{Text/Story.tex}

\section{Discussion}
\vspace{-1mm}
\label{sec:discussion}
\input{Text/NewDiscussion.tex}

\section{Conclusion}
\vspace{-1mm}
\label{sec:limit+future}
\input{Text/limitationsfuturework.tex}
\balance
\Urlmuskip=2mu plus 0mu\relax
\bibliographystyle{plainnat}
\bibliography{caseStudy}

\appendix

\end{document}

%% file: Text/Abstract.tex
System administrators, similar to end users, may delay or avoid software patches, also known as updates, despite the impact their timely application can have on system security. These admins are responsible for large, complex, amalgamated systems and must balance the security related needs of their organizations, which would benefit from the patch, with the need to ensure that systems must continue to run unimpeded. In this paper, we present a case study which follows the online life-cycle of a pair of Microsoft patches. We find that communities of sysadmins have evolved sophisticated mechanisms to perform risk assessments that are centred around collecting, synthesizing, and generating  information on patches. These communities span different Virtual Communities of Practice, as well as influencers who monitor and report on the impact of new patches. As information is propagated and aggregated across blogs, forums, web sites, and mailing lists, eventually resulting in a consensus around the risk of a patch. Our findings highlight the role that these communities play in informing risk management decisions: Patch information is not static, and it transforms as communities collaborate to understand patch issues.


%% file: Text/Introduction.tex

Timely installation of patches (updates) on systems is vitally important to protecting them from attacks but the decision of when or if to patch also involves an analysis of the risks the patch poses. Unfortunately those risks are rarely well understood at the time the patch is released and are instead slowly discovered by the the patching community as the patch is installed on a wide variety of systems~\cite{shostack2003quantifying}. Prior interview research with administrators suggests that they use a variety of online sources to inform their patching decisions~\cite{li2019keepers, tiefenau2020security, jenkins2020eurosp, martiusdoes}. This work expands on this observation by taking an in-depth look at two Microsoft patches that at time of release had no known issues and then later problems were identified from community reports resulting in multiple Hotfix releases and changes to official guidance. The paper uses a case study approach~\cite{YinRobertK2014Csr} to understand the types of information available to admins at different points in time and how that information evolves over time both because of updates on official channels and ongoing conversations in online communities. 

Patching software is a challenging problem for admins due to a combination of a lack of information and the risks associated with both installing and not installing a patch~\cite{li2019keepers, tiefenau2020security, martiusdoes}. Patches often address security issues in the software so choosing not not patch risks those security issues being exploited by attackers~\cite{shostack2003quantifying}. WannaCry is a good example of this risk as the attackers exploited a security vulnerability for which a patch had been available for months~\cite{lessonslearned}. So not patching means risking these types of attacks. But installing a patch is not a risk free proposition either. By definition patches make changes and those changes can impact systems in unexpected ways especially for organizations that are running uncommon setups. For example, a patch inadvertently canceled food stamps for a large number of Californian residents~\cite{foodStampCancel}. Or more seriously, in 2018 a Microsoft patch started deleting files from home directories~\cite{bright_2018}. So patching risks system downtime or other more serious problems. 

Time is also a crucial factor when patching and the risks of installing vs not installing change as time progresses~\cite{moore2002:CodeRed}. When a patch is first released, very little is known about it. The organization that released it may have done some testing but that testing is likely limited to common setups and will not include all possible variations of systems or dependent software. So the risk of the patch causing problems is highest at the time of release. As time progresses, issues with the patch become apparent, users complain about problems, work around are posted, and the organization behind the patch may even release a fixed version. In other words, the longer a patch is out the more likely it is to be problem free or at least to have well documented problems~\cite{beattie2002}. The inverse is true for security vulnerabilities. When a patch is first released there are usually few, if any, active exploits in the wild that target the vulnerabilities it patches~\cite{bilge2012before}. But as time progresses attackers become aware of the vulnerability and exploit it. Hence, the longer an admin waits to patch software the greater the risk of security compromise. 

In other words, at the time of patch release there is no real way for an admin to predict if it will cause problems or what future attacks it would prevent. To manage these risks, admins turn to online information sources to decide things like which patches to prioritize, what patches to delay installing, and to find information about how to troubleshoot or apply work arounds for patches that are not working as desired~\cite{jenkins2020eurosp}.
Ideally this online investigation is also done alongside testing the patches on a test environment or staged roll out where the patch is first deployed to low risk systems. Though not all admins have such infrastructure options~\cite{li2019keepers, tiefenau2020security}. 

In this work we look at the information available to admins from a variety of sources and how that information evolves over time. More specifically, we focus on the following two research questions:

\begin{itemize}
    \item[RQ1] How does the knowledge of risks associated with a patch develop over time?
    \item[RQ2] How do the various communities support the collation and synthesizing of available information for sysadmins and other patching communities?
\end{itemize}

To do so, we conduct a case study of two Microsoft patches, both of which were marked security critical at the time of their release but later were found to have significant enough problems that Microsoft released a Hotfix followed by a second Hotfix to address security issues introduced by the first Hotfix. While this case is not necessarily typical, it does represent the type of situation admins must continuously be concerned about: a critical security patch that is initially reported to be safe to install but is later found to cause problems. The case also exhibits several features that occur somewhat frequently though rarely all in one case: community generated patch prioritization suggestions, reports of possible problems, community diagnosis of the problems, challenges around communication with Microsoft, release of Hotfixes, and dealing with official documentation changing. 



We show that the online communities around sharing patch information are active and provide admins with a range of information that definitely evolves over time. We also highlight how the types of information provided change over time starting with the official posts from Microsoft, then the patch summaries posted by trusted groups, followed by many posts on web forums discussing patching issues, these smaller posts are then summarized by respected community members to present an updated understanding of patch risks and work arounds. In the case of problems, Microsoft or other trusted groups may also post updates. The case highlights the fluid nature of information and how new information is generated and summarized in a community-wide effort which makes it possible for admins to follow along with individual posts or just follow trusted people to get regular summaries, or even only follow Microsoft to get the big points. 


%% file: Text/Literature.tex

\subsection{Sysadmins' work practices}
The seminal works on sysadmins and their work practices are a collection of  ethnographic field studies \cite{barrett2004field, haber2007security, haber2007design, bailey2007activity, barrett2004people, barrett2005usable, kandogan2005security, maglio2003distributed, haber2011collaboration}, which was later combined into a book~\cite{kandogan2012taming}. 
This work highlighted that the work undertaken often involves complex problem solving where many people and data sources must be consulted to conduct the work. Consequently, they also observed that sysadmin work is highly collaborative in nature with admins regularly needing to communicate with others and often using multiple channels to do so~\cite{haber2011collaboration, maglio2003distributed, haber2007security, kandogan2012taming}. 
The work also identified subtle but distinct differences between the work practices of security administrators and other sysadmins \cite{haber2007security, kandogan2005security}, as security work is often event driven and responsive in nature. 

\subsubsection*{Sysadmins' information seeking and requirements}
Sysadmins value accurate and reliable information sources. 
Velasquez et al. label the role of sysadmins as a \emph{broker technician}~\cite{velasquez2009system}. 
Sysadmins must maintain systems to ensure end-user satisfaction, while sustaining a relationship with the technical community through knowledge sharing. 
Velasquez et al. found that sysadmins valued specific attributes of information. For example, accuracy and verification of information were considered to be important, with credibility and reliability features impacting perceived system quality~\cite{velasquez2008work, velasquez2008designing}. Furthermore, task complexity was found to be a significant indicator of sysadmins' need to verify information~\cite{velasquez2008sysadmins}. 

Hrebec and Stiber~\cite{hrebec2001survey} highlighted the importance of \emph{situational awareness} in sysadmins' work. They surveyed 54 sysadmins and found that they estimated their overall knowledge of their systems to be 77\%, indicating the convoluted nature of the systems they dealt with. When asked to attribute the source of this knowledge, almost half (49\%) stated it was hands-on experience. 

\subsection{Sysadmins, Security, \& Patch Management}
Recent work on sysadmin practice has tended to focus on configuration errors~\cite{xu2015systems, xu2016hci, dietrich2018investigating}, with more detailed investigations on SSL~\cite{fahl2014eve}, HTTPS deployment~\cite{krombholz2017have, tiefenau2019usability}, and firewalls~\cite{voronkov2019}. 

The literature on sysadmins touches upon instances of patch management~\cite{kandogan2012taming, velasquez2008work}. 
Shostack \cite{shostack2003quantifying}  illuminated the complexities of patch management and the sheer scale of work required from sysadmins as they balanced the business needs against the security requirements of the organizations they represented. 
Beattie et al. \cite{beattie2002} elaborated on Shostack's work through the construction and evaluation of a model to identify the optimal time to apply an update. Their results found two optimums, one at 10 days and one  at 30 days following the patch's original release. 

Li et al.~\cite{li2019keepers} provided the results of a survey of 102 sysadmins, with a complimentary sample of 17 in-depth interviews. The study's findings detail the  patching process used by sysadmins. 
The stages found are broadly similar to the update stages identified by previous research with end-users~\cite{vaniea2016}. Notable differences include the reliance on testing updates in pre-deployment state to ascertain the risk of a patch to systems. The work also details the breadth of information sources used by sysadmins when identifying a patch exists. Security advisories (78\%), direct vendor notifications (71\%), professional mailing lists (53\%), and online forums (52\%), were the top reported avenues for acquiring update details. These results were similar to those found by Tiefenau et al.~\cite{tiefenau2020security}, who independently conducted a similar study to Li et al. in a European context. 

Additionally, Martius and Tiefenau~\cite{martiusdoes}, used patch release notes to design a survey to identify what information sysadmins reported as being important for their patching decisions. The results showed that 68\% of sysadmins found the lack of patch information made the task of patching more difficult, and that \emph{Known Issues} regarding a patch were highly valued. 

\subsection{Community of Practice \& Distributed Cognition}

Jenkins et al.~\cite{jenkins2020eurosp} explored the PatchManagement professional email list, they performed a qualitative content analysis and found a prominent focus on sharing information regarding the errors caused by patching, and requesting troubleshooting guidance for experienced errors. Similar to previous research~\cite{kandogan2012taming}, the authors identified the mailing list as an example of a \emph{Community of Practice }(CoP)~\cite{wenger1999communities}, a social learning theory which states that humans learn through participation within tight-knit groups with a common goal of gaining knowledge on a domain or area. Through engagement with other members of a community, individuals begin to build upon their understanding of the practice, which can form organically, or be created within an organization~\cite{gunawardena2009theoretical}. \emph{Virtual Communities of Practice} (VCoPs) extends the notion of a CoP by ignoring the need to be physically co-located with other community members, and relying on technology to facilitate engagement~\cite{JOHNSON200145, dube2005impact}.

An additional social learning theory which has been applied to System Administrators and their working practice is that of \emph{Distributed Cognition}~\cite{hutchins2000distributed} (DCog), which details the collaborative efforts between multiple agents to form a single cognitive system~\cite{Busby2001}. Many modern organizations are made up of distributed teams, necessitating hand-offs between them for a task to be completed. For example, Maglio et al. applied the distributed cognition framework with joint activity theory to understand a specific problem-solving instance of system administration~\cite{maglio2003distributed}. The issue required one admin to be in contact with several different teams through email, instant messages, telephone, and face-to-face discussions  as they attempted to debug issues with a firewall. Additionally, Botta et al. applied DCog to Information Technology Security Management (ITSM). They found that challenges in ITSM resulted in a break down of DCog, which ultimately resulted in adverse security risks for organizations~\cite{botta2011toward}. 

\blind{
\subsection{Updates, End-users \& their experience}
The struggles of end-users and updates is comparatively well explored, as
end-users are known to willingly skip or delay updates, thus prolonging their system's window of vulnerability \cite{forget2016or, vaniea2014} and are generally unaware of the security protection ``hidden'' within updating\cite{fagan2016they, mathur2018quantifying, vaniea2014, wash2014out,redmiles2016learned}. This is in direct contrast to the advice of IT security experts who routinely cite updating as a recommend piece of security advice~\cite{ion2015no,reeder2017152, khan2012software}.

Research into why this is the case for end-users has found a number of recurring factors, such as work-flow interruptions~\cite{mathur2018quantifying, wash2014out, vitale2017high}, or unwanted interface changes~\cite{bergman2017cognitive, vaniea2016,vaniea2014, forget2016or} with these inconveniences causing negative experiences which heavily influence decisions when faced with a later updating event~\cite{vaniea2014}.

Theorized similarities~\cite{velasquez2008designing}  between sysadmins and end-users appear to be linked through a social factor of updating. External information through  online reviews, and recommendations from others influence their decision process when updating~\cite{fagan2016they, vaniea2016, mathur2018quantifying, tian2015supporting}. Social processes have equally been shown to impact the adoption rates for end-users and other security mechanism~\cite{das2015role}

To alleviate the problems caused by a lack of updating, removing the human from the decision process through automatic and silent updates has shown to have its benefits when reducing windows of vulnerability \cite{dubendorfer2009silent, sarabi2017patch, nappa2015attack}; however doing so may also result in poorly constructed mental models of how system works~\cite{wash2014out}. Additionally, similar to the findings on sysadmins, automation is not necessarily the magic bullet that will solve all the world's woes~\cite{kandogan2012taming, edwards2008security,vaniea2016}, as updates can be damaging as well as security problematic.}


%% file: Text/methods.tex
\blind{This project began with the scraping of the PatcManagement.org
To illustrate many of the observed types of information shared across the mailing list it was decided to focus in on a set of patches. It is believed that using a case study approach and guidance from 
\cite{YinRobertK2014Csr}, we would be able to articulate the Human factors of these online communities surrounding one Patch Tuesday event. }



We use a case study~\cite{YinRobertK2014Csr} approach to gain an in-depth understanding of the types of information and processes sysadmins must deal with when searching for information about patches. By narrowing our focus to a pair of patches, we were able to do an exhaustive search of available information and fully trace when information became available, where it was posted, how it was incorporated, assimilated, referenced, cited, and used by other people in their own posts, and how it ultimately impacted the growing recommended action consensus within different communities. 

A case study also allows us to develop a detailed case-level understanding of our two research questions around the progressive release of new information (RQ1) and the different ways online communities support information collation and synthesis (RQ2). While this understanding will consequently be case-specific, the lessons learned can inform later more general investigations.
\vspace{-5mm}
\subsection{Case selection}

We decided to focus on Microsoft patches for three key reasons. First, their products are a major feature of the corporate world and the associated support communities are large and well developed. Second, Microsoft patches are released on a regular one month cycle that the community is very familiar with communal processes around patch release. Last, Microsoft labels all patches with a unique Knowledge Base (KB) number, which is then logged in the Microsoft Knowledge Base which is a public repository containing articles, relating to products or user-encountered problems. The KB number is heavily used by sysadmins in their online discussions~\cite{jenkins2020eurosp} making such posts easy to accurately associate with a specific patch.

We selected our patch case with the following features, which exemplify many of the issues sysadmins are known to struggle with~\cite{jenkins2020eurosp, li2019keepers, tiefenau2020security}: is security critical;  has undocumented problems; has a temporary workaround; involves interaction with vendor (Microsoft); and addresses a ``typical'' vulnerability. 

\emph{Security critical} patches are reported to have the highest priority by sysadmins~\cite{li2019keepers, tiefenau2020security} since the vulnerabilities they patch can be serious and lead to compromise if not quickly corrected.

\emph{Undocumented problems}, \emph{temporary workarounds}, and \emph{interaction with Microsoft} are all events that can happen during a patch month. All three are future events that are impossible for sysadmins to predict at the time of patch release but can be problematic later. Undocumented problems are the largest issue since they are problems that may appear without warning damage system functionality. Workarounds are inconvenient, but may allow an admin to keep a problematic patch installed rather than uninstalling it so they are valuable to know about. Finally, interactions with Microsoft provide official guidance and corrections but may require manual intervention to install. 

\emph{A ``typical'' vulnerability patch} was targeted for research and practical reasons. While famous vulnerabilities like Meltdown and Spectre can be quite interesting to study, the media frenzy around them also leads to non-standard behavior. Organizations might allocate more resource or allow sysadmins to take more risks than normal. The amount of media attention also increases the number of people posting about the patches and effectively drowns out the voices of the people we most want to study.
Additionally, while some patches become famous in hind-sight, any critical patch has the potential to later develop into a WannaCry level global disaster.
So for this case study we focused on finding a patch case that is relatively unknown to the general public.

To find potential cases we searched the Patchmanagement.org email archive which has previously been shown to be an online Community of Practice focused on sharing patch information~\cite{jenkins2020eurosp}. We used regular expressions to find email threads containing KB numbers and that looked to have heavily discussed problems, yielding 5 potential cases. We then took a more in-depth look at these cases to determine how well they met our other requirements. Including looking at the KB articles, following links in the emails, and searching other communities like TechNet for the KBs. Ultimately we selected KB4034679 and KB4034664 which were released in August 2017.
Two patches were selected because Microsoft releases patches in pairs, with one containing only security changes (KB4034679) and the other containing all fixes (KB4034664).

\blind{
The patches matched all our case requirements and were marked as security critical by Microsoft. They addressed a security problem that had similarities with a previously in-the-wild exploitation. However, installing the patches could cause visual glitches in a second monitor, which was not officially documented at the time of patch release. The community was able to identify a work-around. The issues also lead to Microsoft releasing new information and fixes to the patch. While highly annoying to sysadmins, the patch issues were never picked up by the popular press, so the patches remain relatively unknown.}

\subsection{Data collection and Analysis}
To collect data we took an approach similar to digital ethnography~\cite{sharp2016role}. Our goal was to collect the information about our patches that sysadmins were actively looking at and engaged with. So we started with the PatchManagement.org email list and searched for all occurrences of the two KBs. We then looked through those email threads and followed all the links in the posts. If the resulting web page was related to the KBs, we recorded it, and then followed all the shared links on that web page. We continued this activity till all such links had been followed. We then took a look at the different sites this activity lead us to and created a list of those that were clearly aimed at sysadmins and patching, including AskWoody and TechNet. We then searched these sites for the two KBs and followed the same process as for Patchmanagement.org. The result was 42 web pages (URLs) that contained content related to our KBs, several of these were forum threads which collectively contained a total of 489 posts or comments.


For data analysis, we chose to focus on the chronology of events as our interests are in how knowledge of risks associated with the patches develops over time (RQ1). Prior work has also shown the patching process to be very time dependent and sequential in nature \cite{li2019keepers, tiefenau2020security, vaniea2016, jenkins2020eurosp}.
The lead researcher read through all 42 web pages including all the posts on them to construct a chronology of events, particularly instances where new knowledge was posted, shared onto a new community, or collated by a member in a larger post or email. 
They then walked through the chronology with another researcher discussing the evidence and any points of uncertainty. 
Both researchers then worked together to group the detailed events into higher level conceptual groups using the previously identified stages of patch management~\cite{li2019keepers, vaniea2016, tiefenau2020security} as a guide. 

\subsection{Limitations}

We specifically selected a pair of patches that exhibited multiple types of problems that admins must handle including: rare problems, a work around, a hotfix which also had problems, Microsoft pulling the hotfix, and a second hotfix. While we feel that these patches exemplify common patching situations that sysadmins must handle, we also acknowledge that it is uncommon for one patch to have this many issues associated with it. We provided an early draft of our results to community members for feedback and validation of our observations. 

A second limitation of our work is in how we found the posts and communities to analyze. We specifically started from the PatchManagement list and used the link structure in posts to find new relevant posts. While this approach allowed us to find all the information considered important enough to link to for our patches, it also excluded information that was less likely to be cited or simply not relevant for the patches we examined. Previous work has shown that professional mailing lists~\cite{li2019keepers, tiefenau2020security, martiusdoes}, and this particular community~\cite{jenkins2020eurosp}, are a prominent source of patch information for sysadmins.

%% file: Text/PlacesPlayers.tex
We start by providing some context about the most prominent communities present in our case study. We also briefly detail the patching model used by our chosen vendor, Microsoft Windows. 

\subsection{PatchManagement.org}
Our start point for this project was PatchManagment.org \cite{patchmanagement.org},  a mailing list which describes itself as: ``the industry's first discussion list dedicated to discussing security patch management topics''. It was formally introduced on November 20\textsuperscript{th}, 2003, and is now hosted on Google Groups. The list is intended to be used by sysadmins and security professionals as an information source. 
The platforms and applications up for discussion are not restricted and are allowed to cover many vendors. It should also be noted that vendors are not discouraged from participation on the mailing list, but are required to announce their affiliations upon joining. The list provides a number of identified suitable topics for discussion, which include users' experience with patches, and notifications from vendors of new or re-released patches.
Notably, the list only allows discussion of vulnerabilities if a patch or workaround exists to mitigate them.

To get a sense of the members we downloaded the list archive consisting of 63,536 emails from 6210 unique email addresses,
with an average number of 10 emails per sender. 
To get a sense of the breadth of backgrounds, we took each of these email addresses and ran them through FortiGuard web filter~\cite{fortiguard}, producing a categorisation of the domains which were active.
The largest categories found were Business (n=168), Private (n=120), Information Technology (n=113), and  Education (n=78). Other industries include Finance and Banking (n=48), Government and Legal Organizations (n=48), and even the Armed Forces (n=6). 
These findings are similar to the list demographics observed by Jenkins et al.~\cite{jenkins2020eurosp}.

\subsection{AskWoody.com}
The AskWoody forum, which is sometimes known as ``The Wailing Wall of IT'', was started in 2004 by Woody Leonhard, as a place for news and discussions surrounding Windows and Office. The site itself, has multiple features related to patching, including designated Master Patch Lists and forum threads for specific platforms. It also features the Microsoft Patch Defense Condition Level~\cite{askwoody_defcon}, or MS-DEFCON system for short, which is similar in nature to the DEFCON system used by the US army. The MS-DEFCON system is a quick reference guide for community members on the current state of released MS patches, giving them a quick way to decide when it is safe to patch, with the scale ranging from "Patches causing issues, avoid deploying" to "All clear, apply patches."

\blind{
\begin{description}
    \item[MS-DEFCON 1:] Current Microsoft patches are causing havoc. Don't patch.

    \item[MS-DEFCON 2:] Patch reliability is unclear. Unless you have an immediate, pressing need to install a specific patch, don't do it.

    \item[MS-DEFCON 3:] Patch reliability is unclear, but widespread attacks make patching prudent. Go ahead and patch, but watch out for potential problems.

    \item[MS-DEFCON 4:] There are isolated problems with current patches, but they are well-known and documented here. Check this site to see if you're affected and if things look OK, go ahead and patch.

    \item[MS-DEFCON 5:] All's clear. Patch while it's safe.
\end{description}
}

A notable feature of AskWoody is the use of the terms Groups A, B, and W~\cite{askWoodyGroupA} to describe people with different patching philosophies loosely linked to their willingness to trust Microsoft. 
The group terms were suggested in reaction to the shift by Microsoft from individual patches to cumulative patches as well as
some patching disasters that were a result of  Windows 10 auto upgrades or  that led to installation of ``tracking'' elements into the OS~\cite{comp_force}. 

\emph{Group A} people regularly install the cumulative updates. They may skip an update due to bugs or issues with the patch, but their philosophy is that patching is important and that new features are wanted. 
\emph{Group B} people regularly install security-only updates, which are not cumulative. They value stability and do not want to install extra features which might bring unwanted elements into their systems. Similar to Group A, they may skip a buggy patch even if it is security related. 
\emph{Group W} people do not believe in patching at all and generally advocate avoiding patch installation whenever possible. This group's philosophy is not recommended by the community, but is respected as existing. 
In our case, Group A would install the cumulative KB4034664, Group B would install the security-only KB4034679 and Group W would install nothing. 

On AskWoody, the different groups are represented through the use of tags with the group name, or in text as a convenient way of referring to a patching approach.  For example, recommendations may state  that a patch is suitable for Group A to install, but that Group B might want to hold off because it adds some extra features. 


\blind{
\begin{figure}[!]
\centering
    \includegraphics[width=0.75\textwidth]{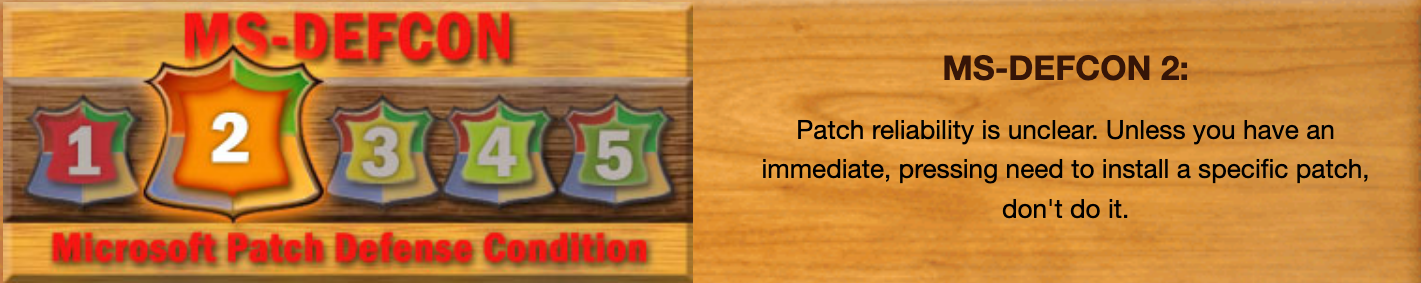}
\caption{Microsoft Patch Defense Condition Level (MS-DEFCON)}
\label{fig:DEFCON}
\end{figure}}


\subsection{Other Communities \& Contributors}

While our discussion focuses mainly on Patchmanagement.org and AskWoody, there are a number of other important partners in the Network of Practice, as we saw many users migrating between forums to discuss or declare issues once patches have been pushed. While linked, we believe that the behaviour exhibited on each respective site and forum is different. For example, the posts which are found on \emph{Technet}, and its sister site \emph{Answers}, appear to be more focused on getting Microsoft's attention, in comparison to the detective work found on AskWoody. 

\emph{G\"{u}nter Born} is a freelance journalist and blogger whose work is captured in this case study. Born began as an engineer, before slowly transitioning into writing and blogging. He writes on a number of subjects, predominantly focused on technology, ranging from books for children and computing, to those dedicated for elderly users. He continues to post on the topic of Windows through his blog, ``Born's IT and Windows Blog''. 
Winner of the Microsoft Community Contributor Award (MCC) in 2011, he is also regularly recognized as a Microsoft  MVP. He can also be found on a number of communities, such as AskWoody, and is a voluntary moderator and wiki editor for the Microsoft Answers forum. 
Microsoft hosts its own official web platform for the discussion of IT related products and issues, called \emph{TechNet}. The site's main function is to provide an area for IT professionals to gain access to information, documentation and have discussions. It has a number of features, such as its own wiki, and documentation library. It also allows for blogs, which are contributed to by Microsoft employees. 

\label{sec:others}

Several other sites offer patch-related content such as curated lists of available patches and blog posts about vulnerabilities.
These lists are created by a wide variety of organizations including: \emph{Ghacks} \cite{ghackSummary}, \emph{Zero Day Initiative} (ZDI)~\cite{zdiSummary}, and the \emph{SysAdmins, Audit, Network,  Security (SANS) institute}~\cite{sansSummary}. News-oriented sites like \emph{Krebs On Security}~\cite{kerbsOn}, also publish Patch Tuesday overview articles. 



\subsection{Patch Tuesday, Platform, \& Model}

``Patch Tuesday'' is the unofficial term used to describe the second Tuesday of the month, upon which Microsoft releases all updates  available for their platforms and products. Due to the scale and dominance of Microsoft, many other companies, such as Adobe, have aligned their patch releases to that of Microsoft.
This monthly event happens at around 17:00 to 18:00 UTC, with the updates being posted on the Microsoft Download Center, and additional related update sites such as Office Update. 

Our patches relate to Windows 7 and Windows Server 2008 R2 platforms, which at the time of our case were still in their extended support cycles. Official support for these platforms expired in January of 2020. 
As mentioned earlier, patches are released in pairs on Patch Tuesday where admins are expected to install one or the other but not both. 
 The first is the \emph{monthly rollup}, which contains all available patches including patches from past months, both feature and security orientated.  
 The second is the \emph{security only}, which contains only the security fixes for that month. 
The monthly rollups are available through Windows Update, Windows Server Update Services (WSUS), and the Windows Update Catalog. The security only updates are only available through WSUS and the catalog.





%% file: Text/Story.tex
Below we present the two communities' perspectives on our two patches KB4034679 (security-only) \cite{seconly} and KB4034664 (cumulative)~\cite{cumlative} from when the communities first learns about them to when they are considered safe to install. At a high level, this story starts with Patch Tuesday release announcements, followed by community testing which flags problems, detective work to find the nature of the problems, a hotfix released by Microsoft to address the problems, more testing highlighting new problems, more detective work, Microsoft removing the buggy hotfix, Microsoft releasing a new hotfix, more testing, and finally a ``safe to install'' announcement.

\subsection{It's Patch Tuesday!}
\vspace{-1mm}
Our case begins on the evening of August 8\textsuperscript{th}, 2017, 17:00 UTC with the standard ``Patch Tuesday'' announcement of released patches by Microsoft. 
August's collection of patches contained 41 security patches, covering all supported versions of Microsoft Windows. Patches were also released for Microsoft Edge, Internet Explorer, Microsoft SharePoint, Microsoft SQL Server, and Adobe Flash Player. The total number of vulnerabilities addressed in Windows on this Patch Tuesday was 48, with 2 vulnerabilities for Adobe Flash Player. 

For each patch, Microsoft also releases an associated KB article with a standard set of information including \emph{improvements and fixes} addressed by the patch, as well a summary of \emph{Known issues in this update}.  The information is updated as issues crop up and are reported by the community, and includes information about proposed and known workarounds.
Security patches are often associated with the Common Vulnerabilities and Exposure numbers (CVEs), which is a unique identifier for a specific security vulnerability. This information is also provided by Microsoft, but on a separate page called the Security Update Guide \cite{securityUpdateGuide}. The guide provides information about the CVE number, severity, and impact. Our cumulative patch is associated with 14 CVEs and the security-only patch addresses 9 CVEs 

\subsubsection*{Patch awareness.}
\vspace{-1mm}

Both official and community curated, annotated lists of the newly available patches (c.f. Section~\ref{sec:others}) are quickly announced and shared through email lists and forums by community members, with citations and credit often given to the original authors/team. These announcements are important because they help admins prioritize their limited time and resources to focus on the patches that are most critical for their organizations.
While summary information often has high overlap between sites, there are also distinctive attitudes and tones to each. 


The Ghacks website, for example, focuses heavily on providing information about the content of the different patches providing, sometimes lengthy, change-log style descriptions of adaptions.
For our patches, Ghacks only notes ``Security updates to Windows Server, Microsoft JET Database Engine, Windows kernel-mode drivers, Common Log File System Driver, Microsoft Windows Search Component, and Volume Manager Driver''~\cite{ghackSummary}.
Ghacks also provides an Excel spreadsheet for download with information about: product impacted, platform, link to KB article, type (cumulative, security, etc.), severity, type of vulnerability addressed, and CVE numbers. Because platforms, patches and CVE numbers have a many-to-many relationship, the August's Excel table had 712 rows for 35 unique KB numbers (patches), and 49 CVEs. 

In contrast, ZDI's  Patch Tuesday summary focuses on the CVEs being patched, including the number of security related patches, and the severity of the vulnerabilities addressed. For our patches, the author draws attention to two selected remote code execution vulnerabilities. Most notable is CVE-2017-8620, which essentially allows a malicious actor to commandeer a target system through construction of a malicious Server Message Block (SMB). It is also noted that this vulnerability is similar to a previously known bug which was exploited in the wild. The author therefore dubs this vulnerability to be ``by far the most critical bug of the month''. 
Following the in-depth look at particular vulnerabilities, a table of all the CVEs addressed in the patches is provided 

On AskWoody's forum, an admin creates a forum thread regarding August's patches entitled, ``Lots and lots of patches''~\cite{lotsandlotswoody}. The post cites the Ghacks blog summary of the security patches released, and provides a snippet of the executive summary for the community members. Within 10 minutes, a member is quick to share the curated list provided by SANS. 

On PatchManagement.org, a moderator sends an email out~\cite{susanAnnounce}, reminding everyone that it is Patch Tuesday. Within this email, they provide a links to official KB articles from Microsoft. 
This thread also experiences a similar rapid response from the community members, with a user supplying the ZDI collated vulnerabilities list. Again, later that evening we see the SANS list shared on the email thread. 

\blind{
\begin{figure}[!]
\centering
\begin{subfigure}[b]{0.5\textwidth}
        \includegraphics[width=\textwidth]{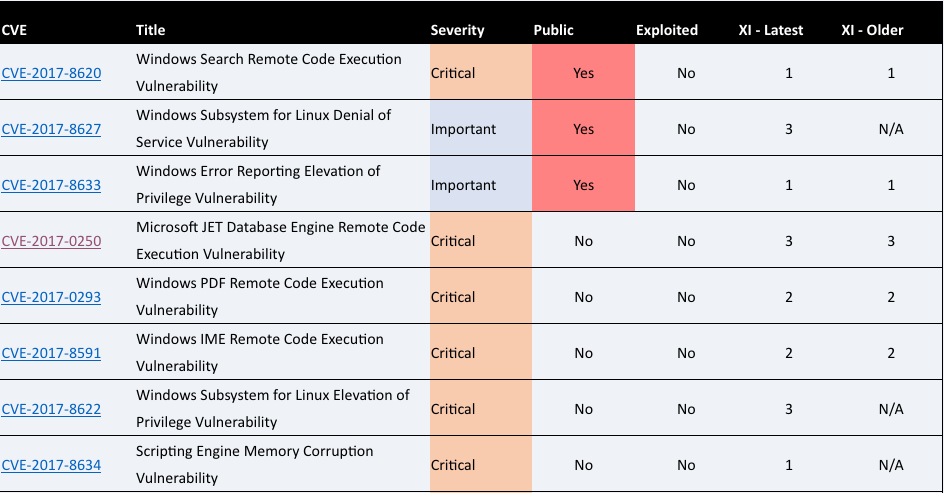}
        \caption{Vulnerability Table}
        \label{fig:zdiTable}
    \end{subfigure}
\begin{subfigure}[b]{0.5\textwidth}
        \includegraphics[width=\textwidth]{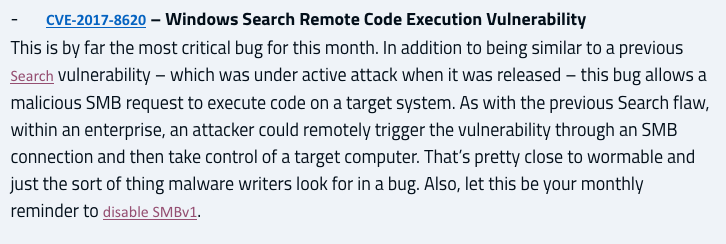}
        \caption{Paragraph on ``most critical vulnerability'' for August}
        \label{fig:zdiCVE}
    \end{subfigure}
\caption{Zero Day Initiative (ZDI) curated patch blog post showing a list of addressed CVEs and a specific warning about the ``most critical bug for this month.''}
\label{fig:ZDI}
\end{figure}

\begin{figure}[!]
\centering
    \includegraphics[width=0.45\textwidth]{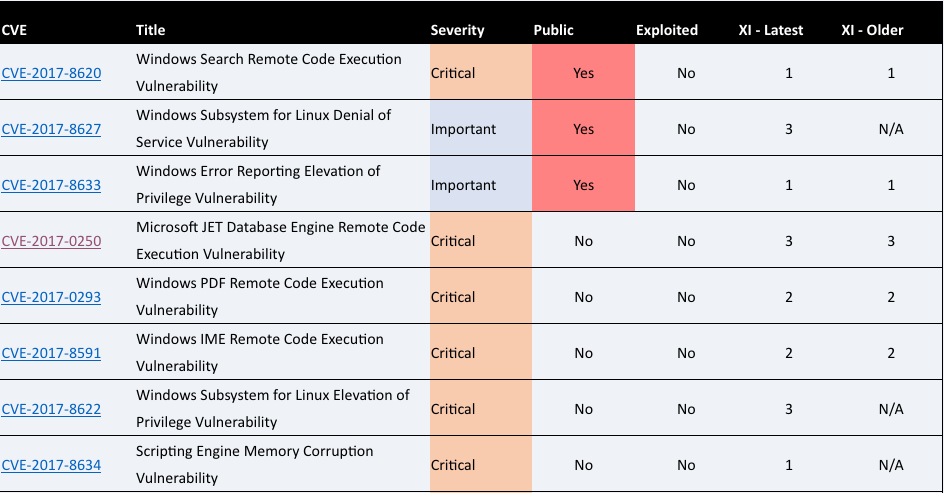}
\caption{Exert of Zero Day Initiative's (ZDI) curated patch blog post showing a list of addressed CVEs.}
\label{fig:ZDITable}
\end{figure}}

\blind{
\begin{figure}[!]
\centering
\begin{subfigure}[b]{0.42\textwidth}
        \includegraphics[width=\textwidth]{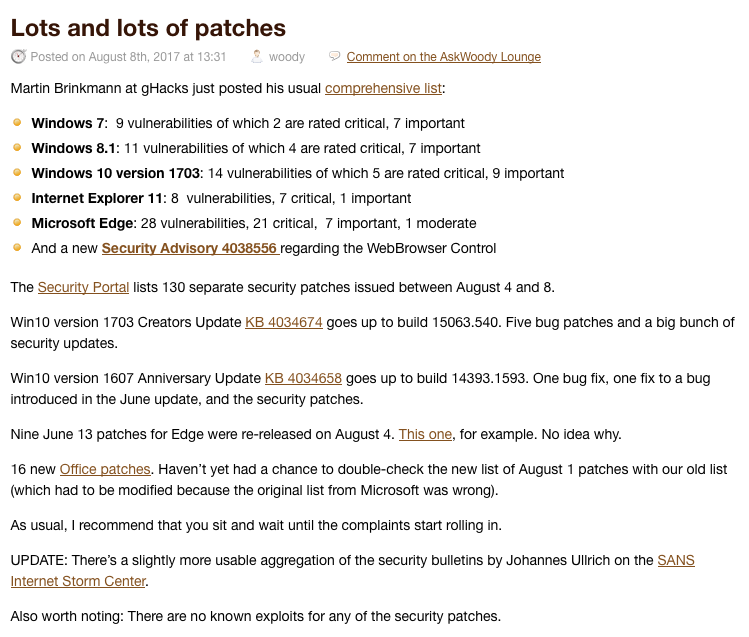}
        \caption{Woody's Forum Post}
        \label{fig:annWoody}
    \end{subfigure}
\caption{Announcements from Woody and PatchManagemnt.org}
\label{fig:Announcements}
\end{figure}
}
\vspace{-1mm}
\subsubsection*{Update identification and prioritization}
One of the first tasks an admin needs to accomplish on Patch Tuesday is to identify and prioritize the patches that they will need to install. The announcements and curated lists described above help  by providing the necessary information in a single source. This  is essential to reduce time spent information seeking, as patch information can be widely dispersed across a number of sources~\cite{li2019keepers}. 

The AskWoody's announcement thread for that month's patches is also quickly used to discuss triage-type issues such as number of patches, the severity, the systems impacted, and the number and type of vulnerabilities. Such centralised information aggregation allows experts to identify and elaborate on patches which are particularly security critical. In essence, this advice informs sysadmins of the most pressing security risks addressed in the patch release. 



\blind{As discussed in the sections above, patching is time-critical. If a CVE is known to be exploited in the wild, then getting any patches associated with it installed quickly is a high priority. But patching also comes with the risk of introducing a problem into the system, so time must be allotted for testing. Since testing all 40+ patches at once is impractical, those resources need to be dedicated to the patches that are both security-critical and impact critical infrastructure. }

Digging into our two patches a bit more, a sysadmin would learn that the patches correct a total of nine unique vulnerabilities, with the cumulative patch correcting an additional five vulnerabilities relating to Internet Explorer. 
As the table shows, they range from moderate to critical in nature, as described by Microsoft themselves. The aforementioned ``most critical bug of the month'' is contained within both of our patches, likely resulting in the patch rising to the top of sysadmins' lists of patches to apply for the respective Windows versions.
\vspace{-1mm}
\subsection{``All code is guilty, until proven innocent''}
\vspace{-1mm}
\subsubsection*{Self reporting initial tests}
One of the well documented reasons sysadmins delay or avoid patching is the risk that the patch may introduce new bugs into the system~\cite{beattie2002,shostack2003quantifying,moore2002:CodeRed}. There are two common ways of mitigating the risk: 1) test the patches to find the ones with problematic errors, 2) wait for others to find the errors. 

Sysadmins who have test environments, and consequently lower risk, install patches in those environments and report findings on forums and email lists. Similarly, sysadmins who perceive either the risk of not patching to be quite high, or the risk of problems quite low, will patch immediately and report the results. Looking at the AskWoody's forum thread~\cite{lotsandlotswoody}, users begin to share initial test results, while stating their patching philosophy (here: Group A), the exact system used, and the exact patch installed: 
\begin{myquote}
    ``Group A, Win 7, SP1, X64 Home Premium  installed  [Cumulative Update KB].  No problems so far. I only use this computer for print, email, internet.''
\end{myquote}
In a more comprehensive post the following day, 9\textsuperscript{th} notes an unexpected change to their settings:
\begin{myquote}
    ``Enabled and started Windows Update on my Win 7 virtual machine. It ran a couple of minutes and reported 2 important and 2 optional updates available$\ldots$ I chose to hide the recurring optional KB2952664 ``telemetry'' update again. 
    
    [Two screenshots of the patching UI.]
    
    The updates went in smoothly, the reboot was clean, no new errors or warnings in the System Event Log. A check for changes: BITS service was changed from DEMAND\_START to AUTO\_START. Further testing is planned.''
\end{myquote}
A later poster asks if the setting changes caused issues, but is told by a third member that the change is unlikely to cause issues and is easily reversed. On this particular thread, no community member identifies potential warning signs regarding our updates.

\subsubsection*{Discovery of errors and issues}

When sysadmins do run into issues, posters tend to report it to places like PatchManagement.org or Woody's Forums by creating a new  thread, with a heading along the lines of: ``I applied patches X and Y happened. My system is running Z etc. Is this happening to anyone else?''~\cite{jenkins2020eurosp}. 
Each thread tends to contain key contextual data, including which patch was installed, what platform it was installed on, and exactly what problem is being observed or discussed. 
The posts are requests to the various communities to help troubleshoot issues.
The bug reporting behaviours appear to happen simultaneously in multiple corners of the Internet. They are also not limited to sysadmins, many end-users also report on the outcome of installed patches, particularly if there are problems that they need help to correct.





For example, one of the earliest indications that our patches have issues is a post to a support forum for IfranView~\cite{ifranerror}, which is a a freeware graphic viewer and editor for Microsoft Windows. Similar to what we see elsewhere, the user creates a new thread, in which they report their issues. Posted on 10.08.2017 at 05:42pm, it states:

\begin{myquote}
``I'd like to report a bug, I have encountered with the 32-bit version 4.37 and the 64-bit version 4.44. It was encountered directly after installing Windows update [Cumulative Update KB], and solved immediately after de-installing said update.

Computer data: OS: Win7 Home Premium SP1, 64 bits, i3 core, 16gb ram, Video Card: AMD Radeon R7 200 Series

The Bug: Going fullscreen on my 2nd monitor causes the images to lag. Meaning if I go to the next image the old image remains on screen. In the case of a slide show, the screen remains black, but images are recovered after dragging an alternate window across it, and in its trail the original image temporarily returns but never fully
...''

\end{myquote}
\vspace{-3mm}
In essence, the application of the cumulative  patch creates a rendering bug on applications displayed on a 2\textsuperscript{nd} monitor,  which, although not debilitating, could easily interrupt or restrict the work-flow of end-users.




\subsection{Understanding Patches and their Errors}
\vspace{-1mm}
\subsubsection*{Synthesising similar bug reports}
Key members of the community keep tabs on patch-related bug reports, even when they are posted on non-patching related forums, like in the IfranView example above. 
Those with many years of experience seem to have become patch soothsayers, picking up on subtle vibrations from the Web, which they then synthesize into analysis blogs and forum posts for consumption by the community. 
These posts consolidate the disparate information which is being posted all across the Internet, and put it into a single comprehensive form. Interestingly, these posts also make heavy use of links and attribution, citing the bug reports that lead to that conclusion and, where appropriate, acknowledging contributors by name.

For our patches, we first see this type of summary post on G\"{u}nter Born's blog~\cite{bornsecondscreen}, on the 12\textsuperscript{th}, 4 days following Patch Tuesday. 
Born says that his interest was peaked through a post on Heise Online's forums~\cite{heisepost}, a German based media company focused on the IT sector. Born provides links to all of his gathered evidence, and in the case of the German language posts, provides translations into English. 
Born's investigative work is impressive in that he finds four other similar issues, and his citations include Matlab forums~\cite{matlabpost}, TechNet~\cite{technetpost}, Nvividia forums~\cite{nvidiapost}, and a singular comment on a news post on Softpedia~\cite{softpedia}. These cited posts are all similar to the IfranView post above, with varying degrees of information provided. The TechNet post in particular provides a large amount of detail supplemented with screenshots of the bug. 

After highlighting the evidence that this is not a one-off event, Born provides known workarounds, which were provided by the users of the Heise forums. The first is to uninstall the guilty patches, which removes the issues, but also results in the system remaining unpatched, and therefore insecure. The second is to enable Desktop Composition, and Born provides a link to an online tutorial~\cite{tutorialdesk} which explains how to do so. 
Although Born posts this on the weekend, the work is noticed by an AskWoody admin, who reacts to the blog by creating a forum thread~\cite{woodyScreenissues}. They begin the thread by immediately citing and thanking Born, before inviting discussion.
\blind{
\begin{figure}
\centering
    \includegraphics[width=.45\textwidth]{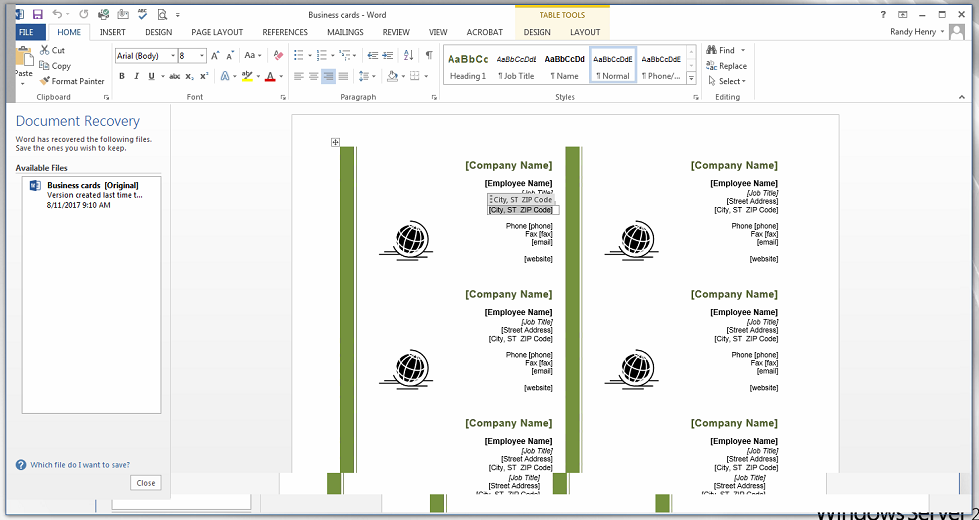}
\caption{TechNet Post showing the rendering issue at the bottom of the image.}
\label{fig:technetPost}
\end{figure}}

\subsubsection*{Proof of concept code}

By the 14\textsuperscript{th}, we see that another community member has provided a link to their blog post~\cite{nineberryblog} on the AskWoody topic thread. Within the personal blog, they detail all known workarounds, which includes upgrading to Windows 10. The important contribution of the post is the inclusion of proof-of-concept code, which replicates the render-bug issue. 
This post is immediately shown appreciation from AskWoody members, and a moderator makes an announcement through their ComputerWorld column~\cite{compworldrenderbug}. 

These posts act as a community wide summation of the current state of knowledge produced from contributions of many differing communities and their members. The work of others is fully acknowledged, with citations, links, and thanks given to all key contributors, including the summation posters themselves. The posts also provide a quick description of the role that screen coordinates play in creating the bug, although the solution advocated by the author is to uninstall both of the offending patches. 

\blind{
\begin{figure}
\centering
    \includegraphics[width=0.43\textwidth]{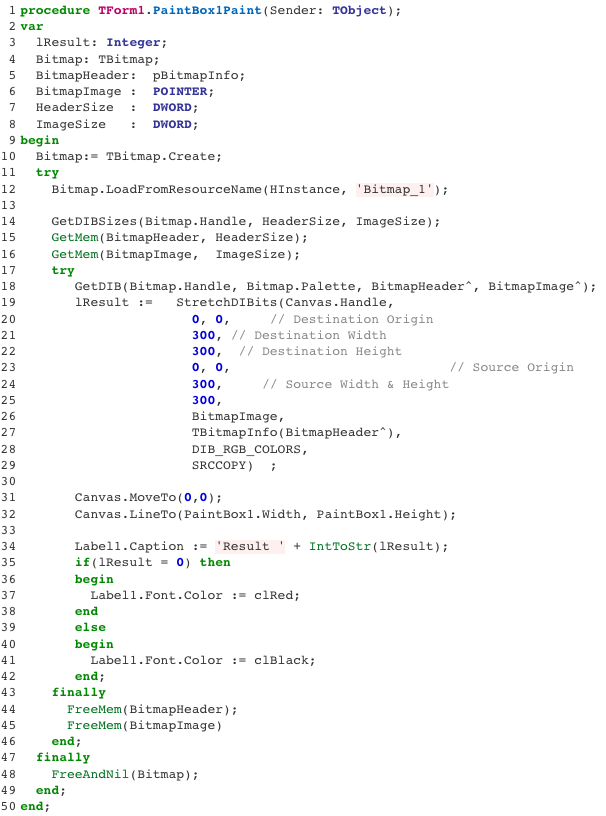}
\caption{Proof of Concept Code}
\label{fig:proofcode}
\end{figure}
}

\subsection{Waiting on Microsoft}
\vspace{-1mm}
Following this summary and background work, a possible option is waiting for an official fix from Microsoft. Enough work and information has been gathered to decide that the issue is related directly to our patches. Therefore, sysadmins and users can now justify waiting for an official response from Microsoft. However this may not be suitable for all, leaving some admins in a bind. Should they push these patches, which are known to cause issues, potentially increasing workload by manually applying workarounds such as screen coordinate adjustment, and impacting the workflow end-users? Or should they avoid patching for now and mitigate through other means? 
Remember, our patches fix a vulnerability which was seen as the most critical of the month, hence a sysadmin's unique context will contribute to their patching decisions

Each of the forum posts continue to have some level of interaction as time passes, mostly confirming that the issue persists for them and providing more data-points and observations. On the TechNet forum post mentioned earlier, which is an official forum and channel to Microsoft, many users indicate that the issues are not unique to the original poster and the amount of users and activity on the post gives an indicator to Microsoft that it must react and address the issue. This is a logical approach for Microsoft, as it has been argued that online reviews and comments could be used to ascertain software quality \cite{Groen2017UsersT}.

\subsection{Hotfix spotted Saturday}
\vspace{-1mm}
Monitoring developments, or lack thereof, is part and parcel for patch management. Given the security implications, sysadmins are expected to be reactionary to any developments and must be able to adapt. Moreover, their work does not finish at the weekend as systems can be attacked at anytime, and malicious agents do not seem to take vacations. Therefore, it is clear that community members must remain aware of related threads and responses from stakeholders involved.

Two weeks later, and 18 days since the original release date,  we see a response from Microsoft in the form of hotfix, KB4039884~\cite{hotfix}. An anonymous poster comments on the AskWoody's thread~\cite{woodyScreenissues}, on the 26\textsuperscript{th}:
\begin{myquote} 
	``Microsoft has acknowledged the bug in their Known Issues for August and releases a patch! (Not yet tested since I could never reproduce the issue...but can collaborate with co-worker on Monday.)''
\end{myquote}

This post happens on a Saturday, but spurs the AskWoody community into action immediately. A separate thread is created~\cite{woodyhotfix}, detailing the hotfix's release, and allowing for space to host detective work. Moreover, Born also notices this and, although slightly behind the AskWoody community, creates a blog post~\cite{bornhotfix} on the following day. 
However, all is not well as by Sunday morning it  becomes apparent that there are issues with this hotfix, with one community member posting: 

\begin{myquote}
``FWIW I installed the said `fix' and then Windows Update came back and told me I had over 20 important updates and a few recommended ones that needed to be installed. Apparently some of the files that are installed are rather old versions (dating back to aug of 2016), just a few are actually newer than 20 aug 17 (according to autoruns). It does change many dll's [Dynamic Link Libraries]. (autoruns'  `compare' facility is great!)''
\end{myquote}

In summary, applying the official hotfix had reverted the system's Dynamic Link Library (DLL) files to previous versions, making them out-of-date and potentially reintroducing previously-fixed vulnerabilities back into the ``patched'' system. This observation was quickly verified by another member of the community, providing further evidence in the form of screenshots. The  solution suggested is to uninstall the patch. It should be noted that all of this activity and detective work takes place before the official documentation appears on Microsoft Support, and  only once the weekend is over do we begin to see the official KB article and related documentation being circulated through PatchManagment.org, through an announcement email similar to those seen at the beginning of this patch cycle~\cite{susanhotfix} on the 28\textsuperscript{th}. 
\blind{
\begin{figure}
\centering
    \includegraphics[width=0.5\textwidth]{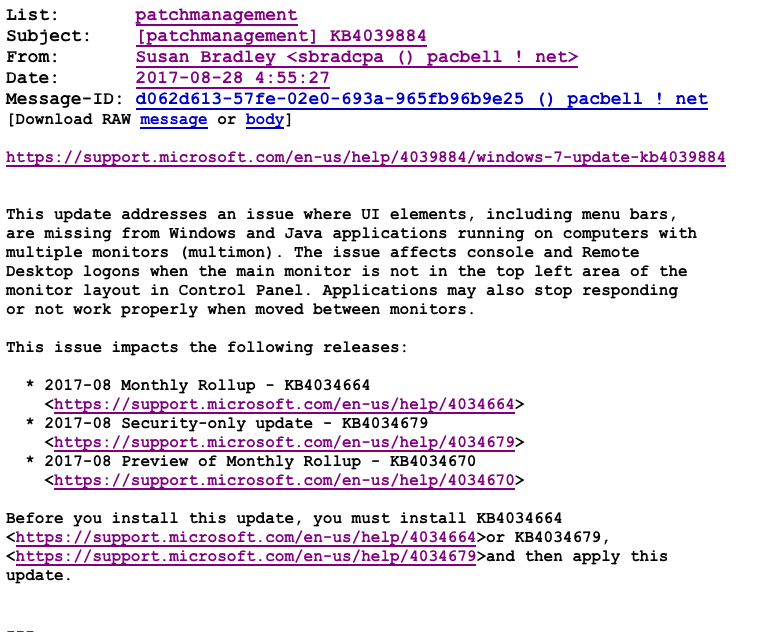}
\caption{Announcement Email of Hotfix Release on PatchManagement.org}
\label{fig:hotfixann}
\end{figure}
}
A moderator summarises the AskWoody community's findings in a column for ComputerWorld~\cite{compworldbuggy}. They are quick to alert readers of the unsafe nature of the hotfix. Using direct quotes from the forum thread created to discuss the hotfix using the post highlighted earlier. 

\subsection{Hotfix Vanishes: ``WTH...''}
\vspace{-1mm}
At some point during the night and into the early hours of the 29\textsuperscript{th}, Microsoft removes the patch from the Update Catalogue, but the KB article remains on Support. Effectively, the associated information regarding the patch remains, however when users attempt to download the patch they are informed that it does not exist. This results in a release of frustration and resentment across all forum and email threads. One example from PatchManagement.org thread on the hotfix release is:

\begin{myquote}
    ``WTH.... I just imported it into WSUS last night and added it to my Software Update Groups.  I guess I'd like to know if it breaks something else now.....''
\end{myquote}



It appears that the unique relationship between a software vendor and sysadmin hinges on a complex form of trust. Sysadmins will find themselves in situations where they must wait for the vendor's response to an issue, so information that is provided is essential in their decision process. Be it data points on uneventful testing, or the detective work on what exactly a patch does, information is king. It is therefore no surprise that the removal of a patch by a vendor without informing sysadmins is tantamount to betrayal. A more detailed response, which epitomizes this dynamic is given by a furious poster on PatchManagement.org:

\begin{myquote}
``Is anyone else disgusted with this policy that Microsoft eventually scrubs its bad 
patches from existence? It is like a great cover-up every time they make a mistake. 
This does not help me as a customer that is looking at a patch wondering if I am 
supposed to be using it to fix a problem. If I did this at work or with my dealings 
with the government I would be fired or in jail. A better and more respectful 
solution would be to state the error, and elaborate on the plan going forward.''

 \blind{Microsoft, perhaps a notice like this would promote goodwill within your customer 
base:

``We regret to inform you there was a problem with the patch KB12345678 issued on 
2017-08-29. We have pulled it from the download repository and will be reviewing 
before it is released again. Please return here for an updated release date. Thank 
you for your patience.''}

\end{myquote}


\subsection{What should we do now?}
\vspace{-1mm}
With the removal of the ``solution'', we now find the community of sysadmins adapting to the new landscape. Given the security critical nature, sysadmins must be responsive, similar to the flexible and reactive nature of their security counterparts~\cite{kandogan2005security, haber2007security}. 
For those sysadmins who were not aware of the hotfix, or were unable to catch it before it vanished, not much has changed. However, for those with access to the hotfix, they can choose to uninstall, again reintroducing the second monitor errors, or stick with the hotfix. As we have discussed, the hotfix itself is potentially security threatening, but one could sink time into updating all the DLL files to remove the reintroduced vulnerabilities. 
The discussion begins to contribute towards which solution is best at the given time. One poster is adamant that the workarounds suggested are not realistic:
\begin{myquote}
``>>The monitor issue is easily resolved by reordering the monitors in the ``normal'' order,  you don't need a patch to \\ >>resolve this problem.

Spoken like a true geek with years of experience. Most home users wouldn't have easily thought of that solution. I know this to be true, because I have done desktop support for a very long time. And by the way, a patch is what broke it; why shouldn't the user expect a patch to fix it?''
\end{myquote}

A moderator of AskWoody also announces the removal of the hotfix through their ComputerWorld Article~\cite{compworldPulled}. This article includes further quotes from forum users that detail findings from independent investigations, again illuminating the problems with the original hotfix. 

\begin{myquote}
``I tested [Hotfix KB Number] in a Windows 7 x64 virtual machine last updated in Sept. 2016.
[...]
It replaces some files with older versions ... for some files [Hotfix KB Number] installs GDR versions of files, replacing existing LDR versions of files. The good news is that uninstalling [Hotfix KB Number] seems to undo these issues, and I recommend doing so if you already installed [Hotfix KB Number]. 
Some of the old file versions that are installed by [Hotfix KB Number] likely have security vulnerabilities that were fixed in newer versions. Thus, installing [Hotfix KB Number] probably exposes your computer to fixed security issues.''
\end{myquote}
The lack of clear communication as to why the patch was pulled left many other online communities in the dark, with the reason only being revealed once the article is shared, as is the case with PatchManagement.org~\cite{susanhotfix} where a member shares the contextual information. Hence, some communities become aware of just what is going on thanks to the work of the communities they are networked with. 

\subsection{``It's baaaaack.''}
\vspace{-1mm}
The Hotfix is then re-released on the 30\textsuperscript{th}, and the communities, again, rapidly notice and inform their respective members. On PatchManagement, a member~\cite{susanhotfix} shares the re-released patch:
\vspace{-1mm}
\begin{myquote}
\blind{https://support.microsoft.com/en-us/help/4039884/windows-7-update-kb4039884}
``This update is now back, now dated August 30. And I'm seeing it in the catalog dated the 30th.

Definitely need a tad more communication.''

\end{myquote}
\vspace{-1mm}
An AskWoody forum thread is created for discussion for this new version of the hotfix~\cite{woodyrerelease}, which quickly descends into further detective work and discussion of Microsoft's communication policy. Many commentators note that the only hint of a previous version from Microsoft is that the KB article contains the instruction:

\begin{myquote}
    ``Before you install this update, you must uninstall any previous version of [Hotfix KB Number]. Then install [Cumulative KB Number] or [Security-only KB Number]KB4034679 before installing this current update of [Hotfix KB Number].''
\end{myquote}

\subsection{Safe to patch $\ldots$ carefully}
\vspace{-1mm}
As discussed in Beattie et al.~\cite{beattie2002}, as time passes, the probability of a patch causing a business essential service to fail reduces, while at the same time, the risk of attack through a vulnerability increases. Therefore, those admins who wish to avoid errors are justified in waiting for the dust to settle. In fact, Beattie et al. showed that the majority of faults in patches are found within the first 30 days, justifying this risk adverse behaviour of waiting to patch. 

Since we now have a working patch, all parties should be satisfied. However, it is not until the 5\textsuperscript{th} of September that we see an ``all-clear'' message from the AskWoody moderator on ComputerWorld~\cite{compworldfinal}. This is four weeks following our chosen Patch Tuesday, and a week before September's official Patch Tuesday. 
The article summarizes the issues of the month, and breaks down update guidance by platform type. An interesting point is the continual focus on addressing those who wish to avoid Microsoft's ``snooping'' and the author's insistence to turn automatic updates off. 


%% file: Text/NewDiscussion.tex
Through our digital ethnography observation of online information about a pair of Microsoft patches we show the different types of information available to admins and how that information evolves over time. We also discuss the role the community plays in both providing information to admins as well as generating information. 

\subsection{Patch Information Evolves}
System administrators consistently report using online information sources to gather information about their systems and to help troubleshoot unexpected system states~\cite{kandogan2012taming,li2019keepers}. Prior work has also shown that they share patching information with each other online~\cite{jenkins2020eurosp}. In this work we dive into how the online information available changes as time progresses. We also highlight that the sources of this information are quite varied and shift over time. Initial information was provided by Microsoft, followed by annotated versions being provided by other trusted groups like Ghacks. But then we see information start flowing from a myriad of untrusted sources, namely people on web forums. This information might be suspect, but instead of ignoring it we see members of the community working to explain, verify, and find workarounds for what is being observed. These observations are then compiled and fed upward via comprehensive posts by more trusted community members till the issues become well enough accepted that Microsoft itself responds by posting a Hotfix to address the issue. 

What is notable here is the location and flow of information that is occurring. With more official sources like Microsoft and trusted organizations providing timely information on patch release, but then being slower to provide updates about potential or confirmed issues with the patches. To fill the information gap, smaller forum- and email-based communities step in to report potential issues as they happen.
It is well accepted that patching quickly is the best approach from a security perspective~\cite{bilge2012before,Verizon2020}. But looking at the reality of online information, it would be extremely difficult for an admin to reliably learn about issues from online sources at the time of patch release or really even a few days later. Our patches required 18 days between patch release and Microsoft officially acknowledging the problem by releasing a Hotfix, which turns out to have security issues itself. So an admin who is following only the official sources might be quite rational information-wise in patching a month behind schedule. Patching any earlier would require a comprehensive testing infrastructure, or the willingness to spend a great deal of time reading forum posts to understand the problems. And even if an admin decided to take the risks and just patch at the time of patch release, they still risk having to spend a great deal of time online to troubleshoot issues and find workarounds as this information exists only in the community curated posts, not the larger official ones.

\subsection{Communities and Networks of Practice}
This case study further highlights the role Virtual Communities of Practice play in patch management through facilitating troubleshooting and issue verification between members. They also provide an informal central repository of the current community-generated status of each patch. 
Communities of practice~\cite{wenger1999communities} traditionally focus on people who interact regularly to learn together about a shared domain, which can range from producing new art to crisis management. Such communities are also found in online virtual spaces~(e.g.~\cite{HaraHew:07,riverin2008sustaining,jenkins2020eurosp}). 

In this case we further observe that patch management information seems to be built on a network of practice, rather than a single community. Many of the key network members, such as Woody, clearly belong to multiple communities and the communities actively reference work from each other. 
This indicates that these Virtual Communities of Practice (VCoP) can be understood as the constituent parts of a larger, loosely connected \emph{Network of Practice} \cite{BrownJohnSeely2002Tslo}. The links within this network are established through fully attributed information that is taken, constructed, reported, and explained uniquely within each respective VCoP. The final judgement, to patch or not to patch, slowly filters through each community to give the entire network much needed insights.

\subsection{Trust in my Community}

Our findings demonstrate the complex trust dynamic which exists in patching between the consumer and the vendor. While VCoP often include representatives of the vendor, our case study has shown why sysadmins may rightly be wary of trusting them. Therefore, they need other trustworthy sources~\cite{redmiles2016think, redmiles2016learned}, accessible through Network of Practice. 

Trust is a  key aspect of sysadmin collaborative work~\cite[p.197-227]{kandogan2012taming}. It is clear from the responses following the pulling of the patch that many sysadmins do not trust Microsoft. This erosion of trust has resulted from past bad experiences which have been discussed and shared within the community, and are archived on their respective sites. The effect of this breakdown of trust echoes the findings of Vaniea et al.~\cite{vaniea2014} and Mathur et al.~\cite{mathur2016they} with end users, who take into account their relationship with a vendor and past bad experiences when updating.

The observed Communities of Practice garner their community member's trust, similar to what others have observed \cite{van1998cooperative}. 
Through access to the experiences and knowledge of others who are familiar with the particular pressures and challenges that sysadmins face, members are mentored and guided through an uncertain patching landscape. Consistent access to useful information makes individual Virtual Communities of Practice credible and trustworthy. This stands in  contrast to  vendors with the lack of consistent patch information and standards~\cite{martiusdoes, abebe2016empirical, aghajani}.

Of particular importance in establishing credibility are  social influencers, or big names and moderators, within these communities. These people build credibility by providing summations of the work of others and provide unsure sysadmins with clear patching targets. This parallels the findings of Das et al. \cite{das2015role}, who showed that social influencers can directly impact the adoption of the security features found on Facebook. Additionally, the work with software developers by Xiao et al.~\cite{xiao2014social} shows that security recommendations from sources with high Internet reputation are viewed as being nearly as trustworthy as that of their peers.

The social nature of these Virtual Communities of Practice facilitates critical discussion of all information.   If a community member contributes incorrect information then the `many eyes' of the community will be able to spot and correct this mistake or misunderstanding, or at least ask for clarification. Therefore the community is quick to correct any knowledge which may be harmful or incorrect, which makes it more trustworthy. Essentially, the advice given, especially when issues have begun to surface, should not be considered as the advice of a single unknown individual, but the collective knowledge of many committed community members all seeking the highest quality guidance and assurances regarding their own patching decisions. 


\subsection{Known Issues are Dynamic}

Our case demonstrates the dynamic nature of patch information, particularly the development of crucial information such as Known Issues~\cite{martiusdoes}. We see that the information sources that sysadmins use are built upon the collaborative effort of community members, who aid in identifying critical updates and share patch quality data. Although our case may not be representative, the case lasted nearly a full patch cycle (4 weeks) which is close to the 30 days found by Beattie et al.~\cite{beattie2002} to be an optimal time to patch. 
One reason patching may take time, is linked to the time it takes for further issues to be identified, investigated, and finally remedied by the vendor. 
Individual communities have evolved unique practices for sharing and generating information, which then spreads through out the network of practice, and pushes vendors to respond. 
Future work should be aim to support community efforts by designing platforms to facilitate this collaborative behaviour in finding and fixing patch issues. Working with these communities can direct attention towards issues which affect the wider patching community reducing potential waiting times. Similar approaches of using user reviews to promote updates to end-users have been investigated~\cite{tian2015supporting} and has potential within the patching context for sysadmins.


%% file: Text/limitationsfuturework.tex
Through a rich case study of two problematic Microsoft Windows patches, we have highlighted the inner workings of the two Virtual Communities of Practice, and that has formed around patch management. We found that patch information is consumed and generated over a period of time, never remaining static as issues, workarounds, and hotfixes are released and new information is generated from community wide testing, and debugging. This information then propagates through a network of communities raising overall patch awareness of the risks of a particular patch.  
Future work should focus on identifying mechanisms to support the generation and flow of information within and between complimentary communities. 